\documentclass[aps,prb,superscriptaddress,twocolumn]{revtex4}
\usepackage{amssymb}
\usepackage{bm}
\usepackage{graphicx}
\usepackage{amsmath}
\usepackage{amsfonts}
\usepackage{color}
\usepackage{ulem}
\usepackage[colorlinks=true,linkcolor=blue,citecolor=blue,urlcolor=blue]{hyperref}

\begin{document}

\title{Topological Micromotion of Floquet Quantum Systems}
\author{Peng Xu}
\affiliation{Institute for Advanced Study, Tsinghua University, Beijing 100084, China}
\author{Wei Zheng}
\email{zw8796@ustc.edu.cn}
\affiliation{Hefei National Laboratory for Physical Sciences at the Microscale and
Department of Modern Physics, University of Science and Technology of China,
Hefei 230026, China}
\affiliation{CAS Center for Excellence in Quantum Information and Quantum Physics,
University of Science and Technology of China, Hefei 230026, China}
\author{Hui Zhai}
\email{hzhai@tsinghua.edu.cn}
\affiliation{Institute for Advanced Study, Tsinghua University, Beijing 100084, China}
\date{\today}

\begin{abstract}
The Floquet Hamiltonian has often been used to describe a time-periodic system. Nevertheless, because the Floquet Hamiltonian
depends on a micro-motion parameter, the Floquet Hamiltonian with a fixed
micro-motion parameter cannot faithfully represent a time-periodic system, which
manifests as the anomalous edge states. Here we show that
an accurate description of a Floquet system requires a set of Hamiltonian
spanning all values of the micro-motion parameter, and this micro-motion
parameter can be viewed as an extra synthetic dimension of the system.
Therefore, we show that a $d$-dimensional Floquet system can be described by a $d+1$-dimensional static Hamiltonian, and the advantage of this representation is that the periodic boundary condition is automatically imposed along the extra-dimension, which enables a straightforward definition of topological invariants. The topological invariant in the $d+1$-dimensional system can ensure a $d-1$-dimensional edge state of the $d$-dimensional
Floquet system. Here we show two examples where the topological
invariant is defined as the three-dimensional Hopf invariant. We highlight
that our scheme of classifying Floquet topology on the micro-motion
space is different from the previous classification of Floquet topology on the time
space.
\end{abstract}

\maketitle

\section{Introduction}

Studying periodically driven quantum systems, which are also referred to as Floquet systems, has become a major research topic in the frontier of quantum
matters~\cite{Bukov2015Universal, Eckardt2015High, Eckardt2017Atomic, Oka2019Floquet}. Periodic driving can be realized, for instance, by illuminating a solid-state material with an electromagnetic wave~\cite{Kimel2005Ultrafast, Fausti2011Light, Mitrano2016Possible, Han2019Time} or by
modulating optical lattices depth~\cite{Gemelke2005Parametric, Lignier2007Dynamical, Sias2008Observation, Parker2013Direct}
or interaction strengths~\cite{Pollack2010Collective, Clark2015Quantum, Clark2017Collective, Clark2018Observation, Behrle2018Higgs, Nguyen2019Parametric} in ultracold atomic gases. Floquet engineering can simulate synthetic gauge fields~\cite{Aidelsburger2011Experimental, Struck2012Tunable, Aidelsburger2013Realization, Struck2013Engineering, Miyake2013Realizing, Atala2014Observation, Greschner2014Density, Clark2018Observation, Gorg2019Realization, Schweizer2019Floquet, Barbiero2019Coupling, Wei2020Floquet, Xu2021Density}%
, and create novel phases such as topologically nontrivial states~\cite{Oka2009Photovoltaic, Lindner2011Floquet, Rechtsman2013Photonic,Zheng2014Floquet, Jotzu2014Experimental, Aidelsburger2015Measuring, Mciver2020Light, Zhang2020Unified} and discrete time crystals~\cite{Else2016Floquet, Yao2017Discrete, Zhang2017Observation, Choi2017Observation, Rovny2018Observation}. It can also realize interesting quantum dynamics such as prethermalization~\cite{Abanin2015Exponentially, Mori2016Rigorous, Abanin2017Rigorous, Abanin2017Effective, Else2017Prethermal, Rovny2018Observation, Rovny2018NMR,Rubio2020Floquet, Beatrez2021Floquet, Peng2021Floquet, Kyprianidis2021Observation} and many-body echo~\cite{Chen2020Many, Lv2020Echoes, Cheng2020Many}.

The Floquet Hamiltonian is a popular tool to describe a periodically driven
system. The key idea of the Floquet Hamiltonian is to effectively describe a
time-periodic system by a time-independent Hamiltonian~\cite{Eckardt2015High, Eckardt2017Atomic, Goldman2014Periodically, Goldman2015Periodically}. Considering a time-periodic
Hamiltonian $\hat{H}(t)$ with $\hat{H}(t)=\hat{H}(t+T)$, we can define a
Floquet effective Hamiltonian $\hat{H}_{\text{\textrm{F}}}$ as ($\hbar =1$)
\begin{equation}
e^{-i\hat{H}_{\text{\textrm{F}}}(\alpha_1)T}=\mathcal{\hat{T}}%
e^{-i\int_{\alpha_1/\omega }^{(2\pi +\alpha_1)/\omega }\hat{H}(t)dt}, \label{Halpha}
\end{equation}%
where $\omega =2\pi /T$, $\mathcal{\hat{T}}$ is the time-ordering operator, and
$\alpha_1/\omega $ is the initial time. Therefore, if an observer
only makes observations at integer periods of time $t=\left( \frac{\alpha_1
}{2\pi }+n\right) T$, this observer cannot distinguish whether the
evolution is governed by $\hat{H}(t)$ or $\hat{H}_{\text{F}}(\alpha_1)$.
In Floquet engineering, one can properly design the driving scheme so that
the Floquet Hamiltonian can display intriguing properties, such as
exhibiting nontrivial topology and novel dynamics.

However, if another observer makes observations at a different set of times $t=\left( \frac{%
\alpha _{2}}{2\pi }+n\right) T$, the corresponding time evolution should be
governed by
\begin{equation}
e^{-i\hat{H}_{\text{\textrm{F}}}(\alpha _{2})T}=\mathcal{\hat{T}}%
e^{-i\int_{\alpha _{2}/\omega }^{(2\pi +\alpha _{2})/\omega }\hat{H}(t)dt}.
\end{equation}%
Now let us use $\hat{U}(\alpha _{2},\alpha _{1})$ to denote an unitary
transformation
\begin{equation}
\hat{U}(\alpha _{2},\alpha _{1})=\mathcal{\hat{T}}e^{-i\int_{\alpha
_{1}/\omega }^{\alpha _{2}/\omega }\hat{H}(t)dt},
\end{equation}%
it is easy to see that
\begin{equation}
e^{-i\hat{H}_{\text{\textrm{F}}}(\alpha _{1})T}=\hat{U}^{\dag }(\alpha
_{2},\alpha _{1})e^{-i\hat{H}_{\text{\textrm{F}}}(\alpha _{2})T}\hat{U}%
(\alpha _{2},\alpha _{1}).
\end{equation}%
Therefore, $\hat{H}_{\text{\textrm{F}}}(\alpha )$ with different $\alpha
$ are equivalent up to a unitary transformation, i.e.,
\begin{equation}
\hat{H}_{\text{\textrm{F}}}(\alpha _{1})=\hat{U}^{\dag }(\alpha _{2},\alpha
_{1})\hat{H}_{\text{\textrm{F}}}(\alpha _{2})\hat{U}(\alpha _{2},\alpha
_{1}).  \label{uni}
\end{equation}%
The $\hat{U}(\alpha _{2},\alpha _{1})$ connects observations at two time
slots within one period $T$, and is also known as the \textit{micro-motion}~%
\cite{Eckardt2015High, Eckardt2017Atomic}. That is to say, although $\hat{H}_{%
\text{\textrm{F}}}(\alpha )$ with different $\alpha $ share the same set of
eigenenergies, their eigenstates differ by an unitary transformation $\hat{U}%
(\alpha _{2},\alpha _{1})$. Hence, the conclusion is that $\hat{H}_{\text{%
\textrm{F}}}(\alpha )$ with a fixed $\alpha $ cannot provide a faithful
representation of this time-periodic system. As a physical manifestation of this
statement, there exists situations that $\hat{H}_{\text{\textrm{F}}%
}(\alpha )$ is topologically trivial but the system exhibits topologically stable edge
states, which are known as the anomalous edge states. Such anomalous edge
states have been explained in terms of the winding numbers of the evolution
operator~\cite{Kitagawa2010Topological, Rudner2013Anomalous, Nathan2015Topological} and higher order topology~\cite{Franca2021Simulating}, and have been experimentally
realized in various systems~\cite{Mukherjee2017Experimental, Maczewsky2017Observation, Wintersperger2020Realization}.

Therefore, a proper description of the anomalous edge states requires complete information beyond $\hat{H}_{\text{\textrm{F}}}(\alpha )$ with a fixed $\alpha$. Previous approaches involve the evolution operator $\hat{U}(t)$ at all time instead of integer periods of time, and $\hat{U}(t)$ is defined as 
\begin{equation}
\hat{U}(t)=\mathcal{\hat{T}}
e^{-i\int_{0}^{t}\hat{H}(t)dt}.
\end{equation}
However, it is easy to see that, although $\hat{H}(t)$ is periodic in $t$, $\hat{U}(t)$ is not. In other word, $\hat{U}(0)\neq \hat{U}(T)$ and the evolution operator is not periodic along the time direction. Hence, an extra operation is designed to impose periodicity in $t$ such that the topological invariant can be well defined ~\cite{Kitagawa2010Topological, Rudner2013Anomalous, Nathan2015Topological}. 

\begin{figure}[t]
\centering
\includegraphics[width=0.45\textwidth]{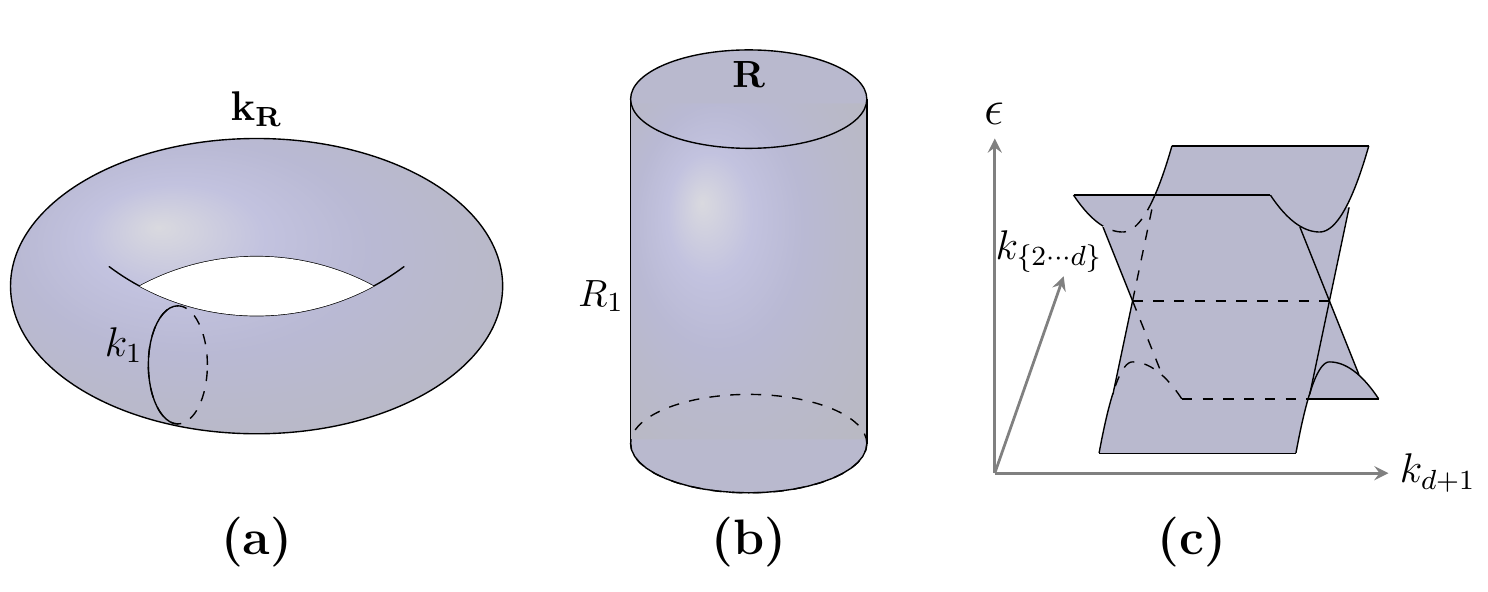}
\caption{(a) The Hamiltonian set $\{\hat{H}_{\text{F}}(k_1,\dots,k_d,\protect \alpha),\protect\alpha\subset [0,2\protect\pi]\}$ can be viewed as a
Hamiltonian in $\hat{H}(k_1,\dots,k_{d},k_{d+1})$ in $(d+1)$-dimension. Here $\mathbf{k}_\textbf{R}$ denotes $\{k_2,\dots,k_{d+1}\}$. (b) A real space
geometry for the $(d+1)$-dimensional system, with open boundary condition in one of the spatial dimension denoted by $R_1$ and periodic boundary
condition in the rest dimensions denoted by $\mathbf{R}=\{R_2,\dots,R_{d+1}\} $. (c) Schematic of energy dispersion for systems shown in (b), as a function of good quantum numbers $k_2,k_3,\dots,k_{d+1}$. This dispersion is flat along $k_{d+1}$ direction. }
\label{band}
\end{figure}

In this work, we propose an alternative scheme that, instead of studying $\hat{H}_{\text{\textrm{F}}}(\alpha
)$ with a fixed $\alpha $ or $\hat{U}(t)$, an effective Hamiltonian set $\{\hat{H}_{\text{%
\textrm{F}}}(\alpha ),\alpha \subset \lbrack 0,2\pi ]\}$ provides complete
information for a time-periodic system. Intuitively, this is because
for a given $\alpha $, the Hamiltonian $\hat{H}_{\text{\textrm{F}}}(\alpha )$
correctly reproduces observations made at $t=(\frac{\alpha }{2\pi }+n)T$. 
Hence, with all $\alpha \subset \lbrack 0,2\pi ]$, observations made at
any time $t$ can be properly captured. Moreover, the advantage of our scheme is that $\hat{H}(\alpha)$ is naturally periodic in terms of the parameter $\alpha$. As one can easily see from Eq. (\ref{Halpha}), $\hat{H}_\text{F}(0)=\hat{H}_\text{F}(2\pi)$. Therefore, the topological invariant can be straightforwardly defined. Here  we will utilize this view to characterize the topology of a non-interacting band, and we will show that
this description can properly capture the anomalous edge states.

\section{General Theory.} 

Let us consider a $d$-dimensional time-periodic
Hamiltonian $\hat{H}(k_{1},\dots ,k_{d},t)$, and the corresponding
effective Hamiltonian set is $\{\hat{H}_{\text{\textrm{F}}}(k_{1},\dots
,k_{d},\alpha ),\alpha \subset \lbrack 0,2\pi ]\}$. Since the effective
Hamiltonian is periodic in $\alpha $, it is therefore quite natural to
consider $\alpha $ as an extra momentum component denoted by $k_{d+1}$. Thus, this Hamiltonian set is replaced by a
Hamiltonian in $\left( d+1\right) $-dimension as $\hat{H}(k_{1},\dots
,k_{d},k_{d+1})$, as shown in Fig. \ref{band}(a). Now we denote $\mathbf{k}_{
\textbf{\textrm{R}}}$ as $\{k_{2},\dots ,k_{d+1}\}$, and their corresponding
real space coordinates are denoted by $\mathbf{R}=\{R_{2},\dots ,R_{d+1}\}$.
When we apply an open boundary condition along $R_{1}$, and keep periodic
boundary condition along other directions, $\mathbf{R}$ spans the surface on
the edge of the system, as shown in Fig. \ref{band}(b). The bulk-edge
correspondence states that, if the Hamiltonian $\hat{H}(k_{1},\dots
,k_{d+1}) $ possesses a nontrivial topological invariant, the system hosts
in-gap surface states localized in the surfaces spanned by $\mathbf{R}$, and
the dispersion of the in-gap states as a function of the good quantum number $%
\mathbf{k}_{\text{\textrm{R}}}$ is schematically shown in Fig. \ref{band}%
(c). A specific feature is that the dispersion is flat along $
k_{d+1} $ direction, since the Hamiltonians with different $k_{d+1}$ (i.e., $%
\alpha $) are equivalent up to unitary transformations. Thus, if such
surface states exist, their dispersion in terms of $\{k_{2},\dots ,k_{d}\}$
should be identical for arbitrary fixed $k_{d+1}$. That is to say, the
Floquet Hamiltonian $\hat{H}_{\text{\textrm{F}}}(\alpha )$ with a fixed $%
\alpha $ also displays in-gap edge states when taking open boundary
condition along $R_{1}$. This discussion shows that the topological
invariant in the $(d+1)$-dimensional Hamiltonian $\hat{H}(k_{1},\dots
,k_{d+1})$ can protect $(d-1)$-dimensional edge states in the $d$
-dimensional time-periodic Hamiltonian $\hat{H}(k_{1},\dots ,k_{d},t)$. Since the physical meaning of the extra dimension comes from the
micro-motion of the Floquet system, we term it as ``\textit{topological micro-motion}''.

Here we should note that the possible topological phase in $\hat{H}%
(k_{1},\dots ,k_{d+1})$ is strongly constrained by the fact that the
Hamiltonians with different $k_{d+1}$ are connected by unitary
transformations and the band dispersion is flat along $k_{d+1}$. This constraint rules out the edge states of $\hat{H}(k_{1},\dots ,k_{d+1})$ being Dirac type. 

\section{Topological Hopf Micro-motion} 

Here we consider a two-dimensional
two-band time-periodic system $\hat{H}(k_{1},k_{2},t)$, and the
Floquet effective Hamiltonian is given by $\hat{H}_{\text{\textrm{F}}%
}(k_{1},k_{2},\alpha )$. Viewing $\alpha $ as $k_{3}$, the eigenstates of the
three-dimensional Hamiltonian are generally written as $|\varphi _{\mathbf{k}%
}\rangle $ with $\hat{H}_{\text{\textrm{F}}}(\mathbf{k})|\varphi _{\mathbf{k}%
}\rangle =\epsilon _{\mathbf{k}}|\varphi _{\mathbf{k}}\rangle $, where $%
\mathbf{k=}(k_{1},k_{2},k_{3})$. We can then introduce a pseudo-spin
direction $\mathbf{n}(\mathbf{k})=\langle \varphi _{\mathbf{k}}|\mathbf{%
\sigma }|\varphi _{\mathbf{k}}\rangle $. Therefore, we define a mapping from the
three-dimensional momentum space $\mathbf{k}$ to the Bloch sphere $\mathbf{n}
$, $f:\mathbf{k\rightarrow n}$. The topology of such a mapping can be
classified by the homotopy group $\pi _{3}\left( S^{2}\right) =Z$, and the
corresponding topological invariant can be described by the Hopf invariant~\cite{Moore2008Topological, Deng2013Hopf, Kennedy2016Topological}.
Considering two different directions in the Bloch sphere denoted by $\mathbf{%
n}_{1}$ and $\mathbf{n}_{2}$, the inverse images $f^{-1}(\mathbf{n}_{1})$
and $f^{-1}(\mathbf{n}_{2})$ are respectively two trajectories in the
three-dimensional momentum space. The Hopf invariant can actually be
described by the linking number of these two trajectories, and this linking
number is independent of the choices of $\mathbf{n}_{1}$ and $\mathbf{n}_{2}$. More details of the definition and the calculation of the Hopf invariant is presented in the Appendix. A nontrivial Hopf invariant can protect edge states in the two-dimensional
surface of a three-dimensional insulator, known as the Hopf insulator~\cite{Moore2008Topological, Deng2013Hopf, Kennedy2016Topological}. With the general theory discussed above, we will show that the Hopf invariant of the three-dimensional Floquet Hamiltonian $\hat{H}_{\text{\textrm{F}}}(k_{1},k_{2},\alpha )$ can also protect one-dimensional edge states in the two-dimensional time-periodic system $\hat{H}(k_{1},k_{2},t)$.

\section{Examples}

Below, we demonstrate this result with two examples. Especially, we will
show that in these two cases, $\hat{H}_{\text{\textrm{F}}}(k_{1},k_{2},%
\alpha )$ with a fixed $\alpha $ is always a topologically trivial
Hamiltonian, but can still process in-gap edge states with open
boundary condition. Thus, the edge states in these cases are the anomalous edge
states. In both cases, we see a definite correlation between the Hopf
invariant in $\hat{H}_{\text{F}}(k_{1},k_{2},k_{3})$ and the presence of
stable edge states in time-periodic system $\hat{H}(k_{1},k_{2},t)$.

\begin{figure}[t]
\centering
\includegraphics[width=\linewidth]{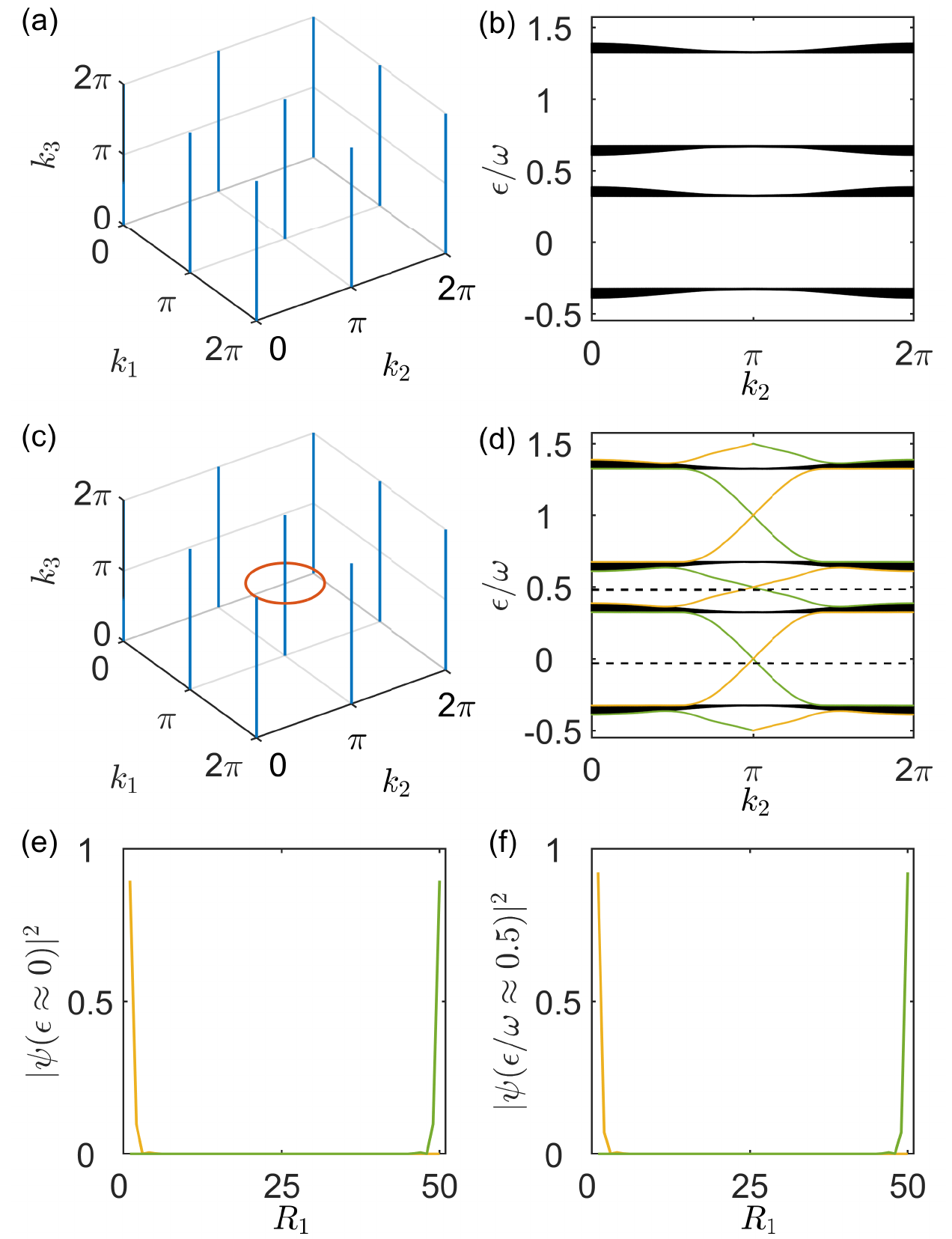}
\caption{Linking number (a, c), spectrum (b, d) and edge states (e, f) of
Model I. (a, c) The inverse images of the south (blue straight lines) and the north (red circle)
poles of the three-dimensional Hamiltonian $\hat{H}_\text{F}(k_1,k_2,k_3)$.
(b, d) the spectrum for two-dimensional Floquet effective Hamiltonian $\hat{H%
}_\text{F}(k_1,k_2,\protect\alpha)$ with a fixed $\protect\alpha$. Here we
have fixed $\protect\mu=-10$ in $\hat{H}_1$ for all plots. We have chosen $%
\protect\mu=-5$ for $\hat{H}_2$ such that $\hat{H}_2$ is a topologically
trivial case in (a) and (b), and $\protect\mu=-2$ such that $\hat{H}_2$ is a
topologically nontrivial case for (c) and (d). (e, f) The real space
distribution of the edge states correspond to the in-gap states shown in
(d), with quasi-energies located at zero-energy (e) and energy $\protect\pi%
/T $ (f) respectively. Here $t_0$ is chosen as $0.1$.}
\label{model1}
\end{figure}

\textbf{Example I.} In the first example, we consider a time-periodic
two-band Hamiltonian
\begin{equation}
\hat{H}=\left\{
\begin{array}{cc}
\hat{H}_{1}, & nT<t\leq nT+t_{0} \\
\hat{H}_{2}, & nT+t_{0}<t\leq (n+1)T%
\end{array}%
\right. .
\end{equation}%
The two-band Hamiltonian can be written as $\mathbf{{h}(k)\cdot {\sigma }}$,
where $h_{x}=\sin k_{x}$, $h_{y}=\sin k_{y}$ and $h_{z}=\mu +\cos
(k_{x})+\cos (k_{y})+\cos (k_{x})\cos (k_{y})$. We take $\hat{H}_{1}=\mathbf{
{h}(k)\cdot {\sigma }}$ with $\mu <-3$ or $\mu >1$, such that $\hat{H}_{1}$
is always topologically trivial. $\hat{H}_{2}$ is chosen as $\epsilon _{0}
\mathbf{{h}(k)\cdot {\sigma }} \mathbf{/|{h}(k)|}$, such that the
band dispersion of $\hat{H}_{2}$ is always flat. For $\hat{H}_{2}$, we can
choose the parameter $\mu $ to make $\hat{H}_{2}$ either topologically
trivial or nontrivial. However, since $\hat{H}_{2}$ has a flat band
dispersion, and by choosing $\epsilon _{0}=\pi /(T-t_{0})$, $\hat{H}_{2}$
always contributes an identity to the evolution operator after one time period. Thus, the effective Hamiltonian is determined by $\hat{H}_{1}$ along, and it is easy to see that, for $\alpha =0$, the Floquet Hamiltonian is always given by $\hat{H}_{\text{\textrm{F}}}(0)=t_{0}\hat{H}_{1}/T$, which is definitely a trivial one. This also means that all $\hat{H}_\text{F}$ are topologically trivial because they are equivalent up to unitary transformations. 

In this model, it can be shown that when $\hat{H}_{2}$ is topologically
trivial or nontrivial, the corresponding $\hat{H}_{\text{\textrm{F}}}(%
\mathbf{k})$ respectively has a zero or non-zero linking number in the
three-dimensional momentum space, and such examples are shown in Fig. \ref%
{model1}(a) and (c). This is because, according to Eq. (\ref{uni}), $\hat{H}_\text{F}(\alpha)$ with different $\alpha$ are connected by a unitary transformation $\hat{U}^{\dag }(\alpha _{2},\alpha _{1})$, and therefore, the eigenstates of $\hat{H}_\text{F}(\alpha)$ with different $\alpha$ are also connected by the same unitary transformation. In this case, it maps the problem to a dynamical quench problem under Hamiltonian either $\hat{H}_1$ or $\hat{H}_2$. It is known from the previous studies of the quench problem that whether a linking number exists depends on whether $\hat{H}_2$ is topologically nontrivial or not \cite{Wang2017Scheme}. 

In Fig. \ref{model1}(b) and (d), we compute the spectrum of the
two-dimensional effective Hamiltonian with open boundary condition along $%
R_{1}$ direction. We can see that when a nontrivial $\hat{H}_{2}$ leads to a
non-zero linking number in the three-dimensional space, the edge states are
present along the one-dimensional edge of the Floquet system. The edge
states are present in both the energy window around zero and around $\pi /T$%
. At the same edge, two edge states at different energies have the same
chirality. Therefore, back-scatterings are forbidden at the same edge, which
ensures the stability of these edge states.

In this model, it is also interesting to ask how the physics recovers the
limit $t_{0}\rightarrow T$. On one hand, as long as $\hat{H}_{2}$ is
nontrivial, the above discussion always results in a non-zero linking
number, which is independent of the choice of $t_{0}$, and this further
leads to the conclusion that the edge states are always present for any $%
0<t_{0}<T$. On the other hand, taking the limit $t_{0}\rightarrow T$, the
Floquet system returns to a time-independent system governed by a
topologically trivial Hamiltonian $\hat{H}_{1}$ and no edge state should
exist. To resolve this paradox, we find that the localization length of the
edge states increases as $t_{0}$ increases. Eventually, when $%
t_{0}\rightarrow T$, states localized at two opposite edges meet in the bulk and gap
out each other.

\begin{figure}[t]
\centering
\includegraphics[width=\linewidth]{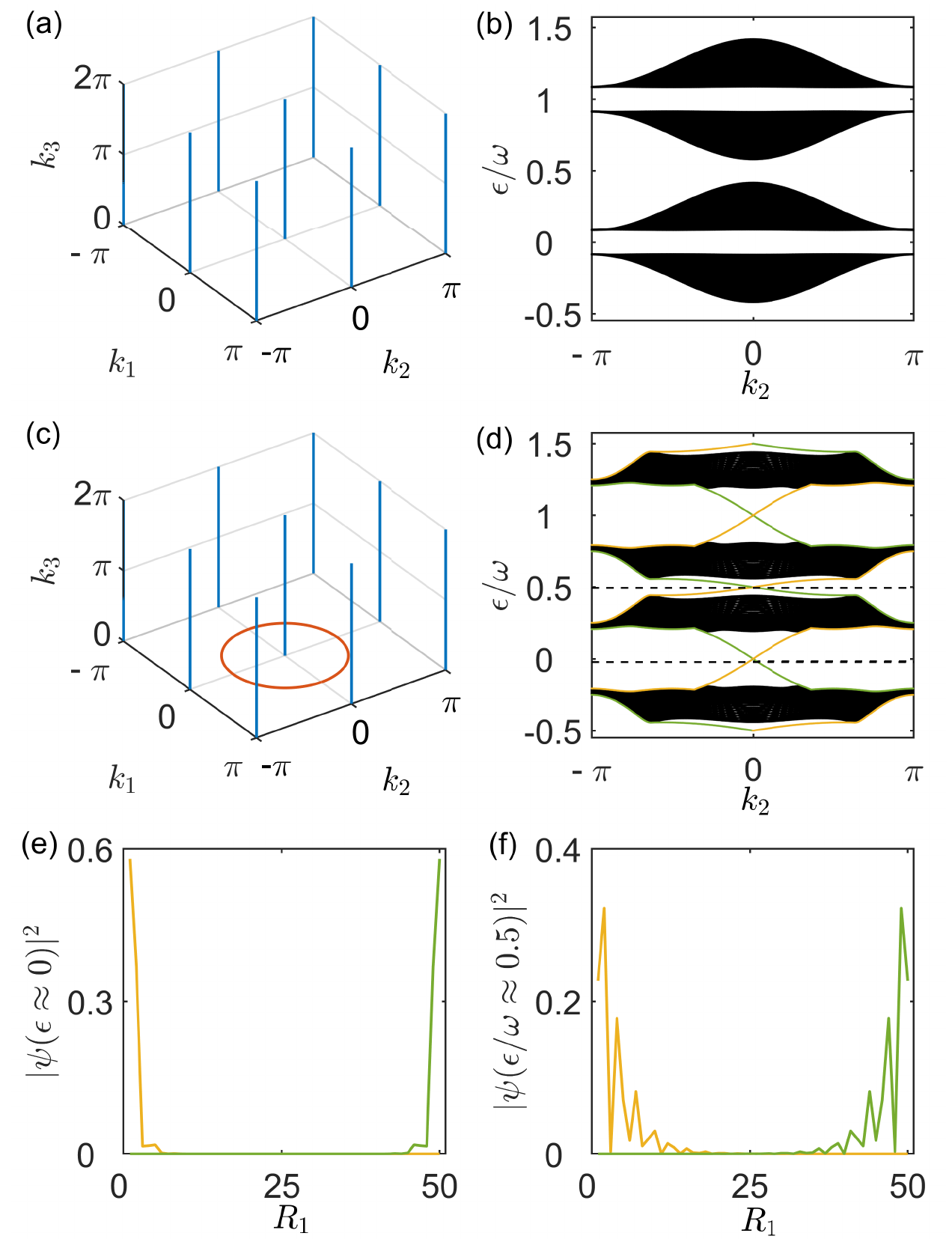}
\caption{Linking number (a, c), spectrum (b, d) and edge states (e, f) of
Model II. (a, c) The inverse images of the south (blue straight lines) and the north (red circle)
poles of the three-dimensional Hamiltonian $\hat{H}_\text{F}(k_1,k_2,k_3)$.
(b, d) the spectrum for two-dimensional Floquet effective Hamiltonian $\hat{H%
}_\text{F}(k_1,k_2,\protect\alpha)$ with a fixed $\protect\alpha$. Here we
have chosen $\protect\mu=-10$ and $\protect\omega=12$ in (a) and (b), and $%
\protect\mu=-2$ and $\protect\omega=4$ for (c) and (d). (e, f) The real
space distribution of the edge states correspond to the in-gap states shown
in (d), with quasi-energies located at zero-energy (e) and energy $\protect%
\pi/T$ (f) respectively.}
\label{modelII}
\end{figure}

\textbf{Example II.} In this example, we consider a time-dependent
Hamiltonian
\begin{equation}
\hat{H}=\mathbf{h}(\mathbf{k})\cdot \mathbf{\sigma }+\sigma _{z}\cos (\omega
t),
\end{equation}%
where $\mathbf{h}(\mathbf{k})$ is the same as described in the Example I and
is time-independent. Here we can also choose different $\mu $ such that $%
\mathbf{h}(\mathbf{k})\cdot \mathbf{\sigma }$ can be either trivial or
nontrivial, and this time-independent part gives rise to two static band
dispersions $\epsilon _{\pm }(\mathbf{k})=\pm |h(\mathbf{k})|$. The $\sigma
_{z}\cos (\omega t)$ term couples the static dispersions to the Floquet
sidebands, which shifts $\epsilon _{\pm }(\mathbf{k})$ by $\pm \omega $ as $%
\epsilon _{\pm }(\mathbf{k})\mp \omega $.

In Fig. \ref{modelII}(a) and (b), we consider the situation that the static
bands are topologically trivial, and we choose a large $\omega $ such that the
static bands do not overlap with the Floquet sidebands. In this case, the
Floquet bands are still topologically trivial and there are no edge states.
In Fig. \ref{modelII}(c) and (d), we consider another situation that the
static bands are topologically nontrivial, and therefore, two bands with
dispersion $\pm |\mathbf{h}(\mathbf{k})|$ have opposite topological numbers.
Then we choose a proper $\omega $ such that a static band (say, band with dispersion $%
|h(\mathbf{k})|$) will overlap with another Floquet sideband (say, band with
dispersion $-|h(\mathbf{k})|+\omega $). In this case, in-gap edge states
occur but the resulting Floquet bands are topologically trivial,
because the mixed two bands originally have opposite topological numbers,
and the band inversion will cancel their topological invariants.

In Fig. \ref{modelII}(a) and (c), we show the linking numbers of $\hat{H}_{
\text{\textrm{F}}}(k_{1},k_{2},k_{3})$. We can see that the linking number
in the three-dimensional space is respectively zero or non-zero for the
situations that the edge states are absent or present. Same as the Example I,
when the edge states are present, they appear in both energy window around
zero and around $\pi /T$ and have the same chirality at the same edge, as
shown in Fig. \ref{modelII}(e) and (f).

\section{Conclusion and Discussion}

In summary, we point out that the
Floquet effective Hamiltonian of a $d$-dimensional system periodically
depends on a micro-motion parameter $\alpha$, and the effective Hamiltonian
set with all $\alpha $ faithfully presents all information of a Floquet
system. Taking $\alpha $ as another synthetic dimension, we view the
effective Hamiltonian set with $\alpha $ as a Hamiltonian defined in $(d+1)$%
-dimension. For a non-interacting band insulator, we show that the
topological number of this $(d+1)$-dimensional Hamiltonian directly protects
stable $(d-1)$-dimensional edge states of the $d$-dimensional Floquet
system~\cite{note1}. Here we would like to highlight again the difference between this
work and the existing works on the Floquet topology~\cite{Kitagawa2010Topological, Rudner2013Anomalous, Nathan2015Topological,Roy2017Periodic, Schuster2019Floquet, Unal2019Hopf, Platero2013Floquet}. The difference
is that here we classify the topology in $\mathbf{k}-\alpha $ space and the
existing works all classify the topology in $\mathbf{k}-t$ space. As
concrete examples, we discuss the situations where 
a three-dimensional Hopf invariant can lead to the 
anomalous edge states. We have explicitly shown two examples and this
theory can also be applied to recent experiments on anomalous Floquet
topological insulator~\cite{Wintersperger2020Realization}, where the anomalous edge states in the experimental models can also be attributed to the Hopf invariant. We note that the Hopf invariant is limited to two-band models, and future works are needed for generalizing to higher band cases. Finally, we expect that this $(d+1)$-dimensional Hamiltonian can also help us to understand other phenomena in Floquet systems such as Floquet discrete time crystal. 

\textit{Acknowledgment.} This work is supported by Beijing Outstanding Young
Scientist Program and NSFC Grant No. 11734010.

\begin{appendix}
\section{The Definition of the Hopf Invariant}

For a two-band model, the Hamiltonian can be written as
\begin{align}
  \hat{H} = \mathbf{{h}(k) \cdot \sigma}.
  \label{Ham}
\end{align}
The groundstate of the Hamiltonian (\ref{Ham}) can be denoted as 
\begin{align}
  \varphi(\mathbf{k}) = \left(
                  \begin{array}{l}
                  \varphi_{1}(\mathbf{k}) \\
                  \varphi_{2}(\mathbf{k})
                  \end{array}
                  \right),
\end{align}
from which we can define a pseudo-spin direction $\mathbf{n}(\mathbf{k}) = \varphi(\mathbf{k})^{\dagger} \mathbf{\sigma} \varphi(\mathbf{k})$. The Hopf invariant of the Hamiltonian (\ref{Ham}) can be evaluated by the integral form ~\cite{Wilczek1983Linking}
\begin{align}
  \text{Hopf} = -\int d^{3} {\bf k} (\bm{j} \cdot \bm{A}),
  \label{Hopf}
\end{align}
where the local current $j^{\mu} = \frac{1}{8 \pi} \epsilon^{\mu \nu \lambda} \mathbf{n} \cdot \left(\partial_{\nu} \mathbf{n} \times \partial_{\lambda} \mathbf{n}\right)$, and $\bm{A}$ satisfies $\nabla \times \bm{A}=\bm{j}$. Note that $A_\mu$ is defined up to the gauge freedom $A_\mu \rightarrow A_\mu - \partial_\mu \Lambda$. Under the gauge choice $\partial^\mu A_\mu = 0$, we have $A_\mu = i \varphi^\dagger \partial_\mu \varphi$. Numerically, we first calculate the ground states $\varphi(\mathbf{k})$ of the effective Floquet Hamiltonian $\hat{H}_{\text{F}}(\mathbf{k} = (k_1, k_2, \alpha))$ at each momentum $\mathbf{k}$, with which $\mathbf{n}(\mathbf{k})$ can be obtained. Secondly, we calculate the local current $j^{\mu}(\mathbf{k})$ and gauge field $A_\mu(\mathbf{k})$. Finally, the Hopf invariant can be obtained directly according to the definition Eq.~(\ref{Hopf}).

\end{appendix}


\end{document}